\documentclass[aps,prl,twocolumn,showpacs]{revtex4-1}  
\usepackage{amsmath}
\usepackage{amssymb}
\usepackage{graphicx}
\usepackage{epstopdf}
\usepackage{suffix}
\usepackage{mathtools}
\begin{document}

\title{Entangling Distant Spin qubits via a Magnetic Domain Wall}

\author{B. Flebus}
\author{Y. Tserkovnyak}
\affiliation{Department of Physics and Astronomy, University of California, Los Angeles, California 90095, USA}

\begin{abstract}
The scalability of quantum networks based on solid-state spin qubits is hampered by the short range of natural spin-spin interactions. Here, we propose a scheme to entangle distant spin qubits via the soft modes  of an antiferromagnetic domain wall (DW). As spin qubits, we focus on quantum impurities (QI's) placed in the vicinity of an insulating antiferromagnetic thin film.
The low-energy modes harbored by the DW are embedded in the antiferromagnetic bulk, whose intrinsic spin-wave dynamics have a gap that can exceed the THz range.  By setting the QI frequency and the temperature well within the bulk gap, we  focus on the dipolar interaction between the QI and two soft modes localized at the DW. One is a  string-like mode  associated with transverse displacements of the DW position, while the dynamics of the other, corresponding to planar rotations of the N\'eel order parameter, constitute a spin superfluid.
By choosing the geometry in which the QI does not couple to the string mode, we use an external magnetic field to control the  gap of the  spin superfluid and the qubit-qubit coupling it engenders. We suggest that a tunable micron-range coherent coupling between qubits can be established using common antiferromagnetic materials.
\end{abstract}

\maketitle

\textit{Introduction.}  The discovery of quantum-impurity (QI) model systems, such as NV color centers~\cite{Doherty2013},  which show long coherence times and  can be initialized and read-out optically~\cite{BarGill2013, Balasubramanian2009,Chu2015,Kennedy2003}, has stimulated a vast interest within the field of quantum computing~\cite{Childress2013,Weber2010,Ladd2010}.  
Direct coherent coupling between single NV centers  has been already observed by several groups~\cite{Gaebel2006, Dolde2013, Bermudez2011}. However, due to its dipolar nature, such spin-spin coupling extends only up to tens of nanometers. 
This distance requirement for their interaction limits the implementation of large-scale quantum entanglement schemes, where the  ability  of addressing each qubit individually must also be preserved. To circumvent this drawback, numerous proposals for the coherent coupling of atomistic qubits 
 revolve around hybrid quantum devices, where distant qubits  interact indirectly via, e.g.,  mechanical resonators \cite{Kolkowitz2012}, superconducting flux qubits~\cite{Marcos2010}, photons \cite{Togan2010}, or spin waves~\cite{Trifunovic2013}.

\begin{figure}[b!]
\centering
\includegraphics[width=1\linewidth]{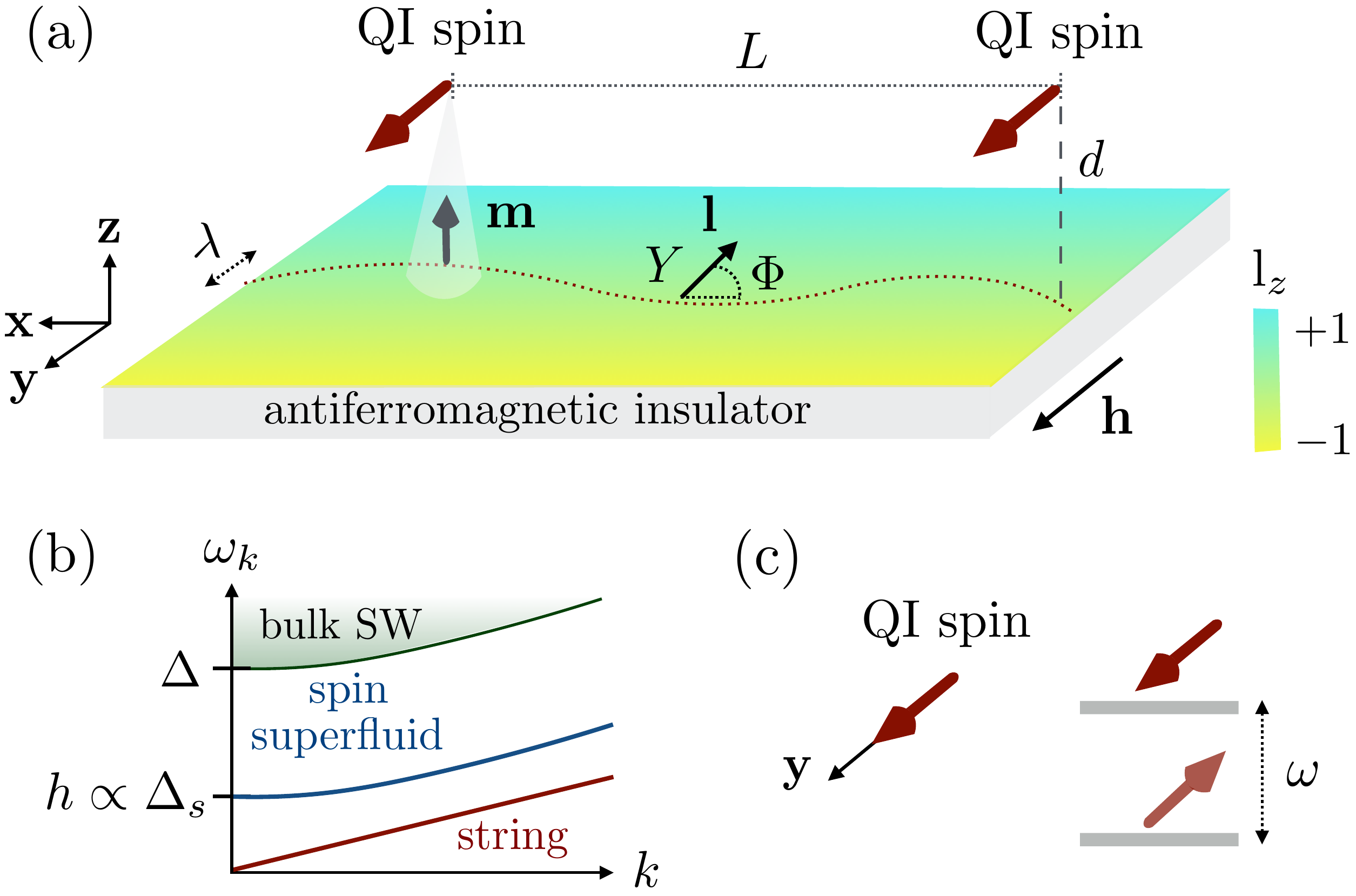}
\caption{(a) The proposed hybrid quantum system: An antiferromagnetic film harbors a DW  of width $\lambda$ along the $x$ axis. This soliton interpolates between the ground states with $\text{l}_{z}=\pm1$  at $y\to\mp\infty$, where $\mathbf{l}$ is the N\'eel order parameter. The $x$-dependent variables $(Y,\Phi)$ identify, respectively, the $y$ position and the azimuthal angle of the order parameter at the DW center. Two QI's,  placed at a height $d$ above the DW and distance $L$ from each other, interact magnetostatically with the film's spin density $\mathbf{m}$. The axial-symmetry breaking magnetic field $\mathbf{h}$ is applied along the $y$ direction. (b) Dispersions of the collective spin modes in the film. Dark green line:  bulk spin-wave (SW) dispersion with energy gap $\Delta$. Blue line: dispersion of the spin-superfluid mode in the presence of the magnetic field $\mathbf{h}=h \mathbf{y}$, which opens a gap $\Delta_{s}\propto h$. Red line: string sound mode. (c) The QI spin $1/2$  is quantized along the $y$ axis, with the level splitting of $\hbar\omega$.}
\label{Fig1}
\end{figure}
 
While the interaction between NV centers and spin waves has recently allowed to probe a range of magnetic phenomena with high  spatiotemporal  resolution~\cite{Casola2018}, hybrid architectures relying on magnetic insulators as building blocks remain relatively unexplored. In Ref.~\cite{Trifunovic2013}, spin waves in microfabricated ferromagnetic waveguides have been proposed to mediate long-distance coupling between  spin qubits.
In this Letter, we instead suggest to employ low-energy excitations associated with extended spin textures, such as domain walls, in an otherwise homogeneous magnetic background. 

Specifically, we consider an antiferromagnetic insulating film with uniaxial anisotropy, which supports an extended domain wall (DW), as depicted in Fig.~\ref{Fig1}(a). The antiferromagnetic DW harbors two types of Goldstone modes associated with its real- and spin-space dynamics~\cite{FlebusOchoa2018}. These are respectively related to the zero modes associated with the DW displacement $Y$ and the azimuthal angle $\Phi$ of the N\'eel order parameter therein. Two QI spins, placed above the DW, interact dipolarly with the collective spin modes of the antiferromagnetic system, whose dispersion relations are shown in Fig.~\ref{Fig1}(b).  By setting the QI resonance frequency within the bulk gap, which is of the order of $0.1\div1$ THz in many of the common antiferromagnetic materials ~\cite{baltz2018}, we can  focus on the excitations endowed by the DW for both  qubit decoherence and qubit-qubit coupling.  The string mode is a translational-symmetry restoring Goldstone mode that cannot be easily gapped (unless we pin the DW position), which may pose problems for controlling the relative importance of the coherent coupling and the decoherence \cite{FlebusOchoa2018}. We show, however, that in the appropriate geometry, the string mode decouples from the spin qubits, at the leading order. 
Therefore, the spin-superfluid mode is left to control both the effective qubit-qubit coupling and qubit decoherence. These can thus be tuned via an in-plane magnetic field, which opens a gap in the spin-superfluid spectrum.
 
In this work, we address two distinct but related problems. First, we consider a spin qubit to couple to a single quantized spin-superfluid mode of a $\sim$ micron-long DW. We show that the associated cooperativity can be large, suggesting that two distant spin qubits can be in principle coupled coherently via a single superfluid mode. Secondly, we look at the interplay between qubit decoherence and qubit-qubit interaction provided by the continuum of modes in an infinite DW.   We find that the spin-superfluid mode can mediate a two-qubit gate with an operation rate of the order of tens of kHz, when the qubits are placed at the distance of a micron from each other. The QI decoherence rate due to the spin-superfluid noise can be found to be  longer (by a factor $\sim 10^2)$ than the gate-operational rate.  

\textit{Main results.} 
We consider  two  spin-$1/2$ QI's with resonance frequency $\omega$, placed at  a distance $L$ from each other and at a height $d$ above the domain wall, as shown in Figs.~\ref{Fig1}(a) and (c).
The QI spin at a position $\vec{r}_{i}$ couples to the stray field $\mathbf{B}(\vec{r}_{i})=  \gamma \int d\vec{r} \; \mathcal{D}(\vec{r},\vec{r}_{i}) \mathbf{m}(\vec{r})$ generated by the antiferromagnetic  spin density $\mathbf{m}(\vec{r})$ via  Zeeman interaction.  Here,  $\gamma$ ($\tilde{\gamma}$) is  the gyromagnetic ratio of the magnetic film (QI spin) and $\mathcal{D}$  the tensorial magnetostatic Green's function \cite{FootnoteD,Guslienko2011}. 
An external  magnetic field $\mathbf{h}=h \mathbf{y}$ sets the QI quantization axis along the $y$ direction and enforces a Bloch domain wall configuration, i.e., $\Phi=0$. As discussed in  details later, these choices lead to vanishing coupling between  each QI spin and the DW string mode. Moreover, the magnetic field opens up a gap $\Delta_{s}=\gamma h$ in the spin-superfluid dispersion, as shown in Fig.~\ref{Fig1}(b). Here, we take the spin-superfluid gap to be much smaller than the spin-wave bulk gap. Thus, at QI resonance frequencies comparable with the spin-superfluid gap, we can neglect the QI coupling with bulk spin waves and focus  on its  interaction with the spin-superfluid mode.

For a DW of length $\ell$, we focus on the coupling between a QI spin and a single  spin-superfluid mode. The latter can be quantized in terms of the magnon creation (annihilation) operator $a^{\dagger}_{k}$ ($a_{k}$) with dispersion $\hbar \omega_{k}$.  The interaction Hamiltonian  can be written as 
\begin{align}
 \mathcal{H}_{\text{int}}= g\, \sigma^{+} a_{k}+ \text{H.c.},
 \label{58}
\end{align} 
where  $\sigma^{\pm}=\sigma_{\tilde{x}} \pm i \sigma_{\tilde{y}}$, with $\sigma_{\tilde{\alpha}}$ being the $\alpha$ Pauli matrix in the QI spin reference frame. Note that, in deriving Eq.~(\ref{58}), we have assumed $g \ll \omega  \simeq \omega_{k}$ .
The cooperativity associated to Eq.~(\ref{58}) can be defined as $C=g^2  \tau_{s}T_2$~\cite{Rabl2010}, where $T_2$ is the intrinsic QI dephasing time and $\tau_{s}$ the DW-mode relaxation time. 
 We find the coupling $g$ as
\begin{align}
g=\frac{ \hbar  \gamma  \tilde{\gamma}}{2}   \sqrt{ \frac{ \lambda \chi \hbar \omega_{k} \left[ \mathcal{D}^2_{xz}(k,d) +\mathcal{D}^2_{zz}(k,d)\right] }{2 \ell}  }\,,
\label{61}
\end{align}
where  $\chi$ is the static uniform transverse spin susceptibility  and $\lambda$
the DW width. Here, $\mathcal{D}_{\alpha \beta}(k,d)$ is the one-dimensional Fourier transform of the magnetostatic Green's function $\mathcal{D}_{\alpha \beta}(\vec{r},\vec{r}_{i})$ at the QI  position $\vec{r}_{i}=(0,0,d)$. This function decreases rapidly as function of the distance $d$ and it is maximized for $k \sim 1/d$.

For a DW of infinite length, we focus on the coupling between the QI spins and the continuum of the spin-superfluid mode. The strength of the effective qubit-qubit coupling and the single-qubit decoherence are parametrized, respectively, by the real ($\chi'$) and imaginary ($\chi''$) part of the spin-superfluid dynamical transverse spin susceptibility $\chi_{zz}(k, \omega)$. We find the coupling as
  \begin{align}
\mathcal{H}_{c}=  \int \frac{dk}{2\pi}  f(k,d) \chi_{zz}'(k,\omega) \cos(k L)\sigma^{+}_{1} \sigma^{-}_{2}+\text{H.c.}\,,
\label{177}
\end{align}
with $f(k,d)= \left[ D^2_{zz}(k,d) + D^2_{xz}(k,d) \right] (\gamma \tilde{\gamma})^2/16$.  In our geometry, the stray field associated with the spin-superfluid mode is transverse to the QI quantization axis. Thus, while  there is no QI dephasing due to purely longitudinal coupling, the spin-superfluid mode gives rise to QI relaxation processes. The associated QI relaxation rate reads as
\begin{align}
T^{-1}_{1}(\omega)&=\coth\left(\beta \hbar \omega/2\right) \int \frac{dk}{\pi} \; f(k,d)  \chi''_{zz}(k,\omega)\,.
\label{68}
\end{align}
 Equation~(\ref{68}) accounts for processes in which the creation or annihilation of a magnon gives rise to a QI transition between its spin states and viceversa. For $\omega < \Delta_{s}$ (or $\omega < \Delta_{s}-1/\tau_{s}$, when accounting for magnetic damping), the magnon spectral density vanishes and the relaxation rate~(\ref{68}) is minimized. On the other side, the real part of the spin susceptibility decays exponentially on the lengthscale $\ell_{s}=c/\sqrt{\Delta_{s}^2 - \omega^2}$, i.e.,  $\chi_{zz}'(x)  \propto   \ell_{s}  e^{-x/\ell_{s}}.$ Thus, to maximize the ratio between the effective qubit-qubit coupling and the single-qubit decoherence, one needs to set the QI frequency just below the gap, i.e., $\omega \lesssim \Delta_{s}$ (or $\omega \lesssim \Delta_{s}-1/\tau_{s}$).

\textit{Antiferromagnetic system.} At temperatures far below the N\'eel temperature, we can describe the low-energy long-wavelength dynamics of a bipartite antiferromagnet in terms of the  directional N\'eel order parameter field $\mathbf{l}(\vec{r},t)$, with $|\mathbf{l}|=1$.
The Lagrangian of an isotropic cubic antiferromagnet with exchange stiffness $A$ and uniaxial anisotropy $K$ can be written as
\begin{align}
\mathcal{L}\big[\mathbf{l}, \dot{\mathbf{l}}\big]&=\frac{\chi}{2}\int d\vec{r} \; \big( \dot{\mathbf{l}}+\gamma \mathbf{l} \times \mathbf{h} \times \mathbf{l} \big)^2 - \mathcal{H}[\mathbf{l}]\,, \; \; \text{with}\nonumber \\
 \mathcal{H}[\mathbf{l}]&= A \big( \vec{\nabla} \mathbf{l} \big)^2 + K |\mathbf{z} \times \mathbf{l}|^2\,.
 \label{92}
\end{align}
Varying Eq.~\eqref{92} with respect to $\mathbf{m}$ leads to the constitutive relation  $\mathbf{m}=\chi\,\mathbf{l}\times\left(\partial_t\mathbf{l}-\gamma\,\mathbf{l}\times\mathbf{h}\right)$ \cite{ Andreev}. Dissipation can be introduced by means of the Rayleigh function $\mathcal{R}[\mathbf{l}]=\alpha s   \int d \vec{r} \; (\partial_{t}\mathbf{l})^2 / 2$, where $\alpha$ is the Gilbert damping constant and $s$ the saturated spin density of both sublattices. 
The model~\eqref{92} admits also a solution for a static  domain wall of width $\lambda=\sqrt{A/K}$. For boundary conditions of the form $l_z(y\rightarrow\pm\infty)=\pm 1$ and using the parametrization $\mathbf{l}=\left(\cos \phi \sin \theta, \sin \phi \sin \theta, \cos \theta \right)$, the DW solution is given by 
\begin{align}
\cos \theta\left(\vec{r}\right)=\tanh\frac{y-Y}{\lambda}, \; \; \phi\left(\vec{r}\right)=\Phi.
\label{DWsol}
\end{align}
By plugging the DW solution~(\ref{DWsol}) into Eq.~(\ref{92}) and promoting the azimuthal angle to a dynamical field $\Phi(x,t)$, we obtain the Lagrangian of the spin-superfluid mode as
 \begin{align}
 \mathcal{L}\big[\Phi, \dot{\Phi} \big] =\lambda \int dx \; \left[  \chi  \dot{\Phi}^2  - A (\partial_{x} \Phi)^2  + \chi ( \Delta_{s} \Phi)^2 \right]\,.
 \label{Lagrangian}
 \end{align}
Following the standard canonical quantization of a harmonic oscillator, Eq.~(\ref{Lagrangian}) can be  quantized in terms of magnon operator $a_{k}$ ($a^{\dagger}_{k}$) with dispersion $\hbar \omega_{k}=\sqrt{(ck)^2 + \Delta^2_{s}} $, where $c=\sqrt{A/\chi}$. From the Lagrangian~(\ref{92}) and the Rayleigh function, we can derive the transverse spin susceptibility of  the spin-superfluid mode as
\begin{align}
\chi_{zz}(k, \omega)= \frac{2 \lambda \chi \omega^{2}}{\omega^{2}_{k}-\omega^{2}- i  \alpha s \omega/  \chi} + 2 \lambda \chi\,.
 \label{susceptibilitySS}
 \end{align}

\textit{Noise.}   The interaction between a QI spin and the antiferromagnetic spin density can be generally recasted in the form
\begin{align}
\mathcal{H}=  \sigma^{+} \otimes X+ \sigma_{\tilde{z}} \otimes Z +  \text{H.c.}\,,
\label{noisyH}
\end{align}
where $X$ and $Z$ are fluctuating fields coupling, respectively, transversely and longitudinally to the QI quantization axis.
The relaxation, $T^{-1}_{1}$, and dephasing, $T^{-1}_{2}$, rates of  each qubit can be written as~\cite{Divicenzo2005}
\begin{align}
 T_{1}^{-1}= \hbar^{-2}  S_{Y}(\omega) \,, \; \; T^{-1}_{2}= \frac{1}{2} T^{-1}_{1} + \hbar^{-2} S_{X}(0)\,.
\label{timesnoise}
\end{align}
Here, $S_{A}(\omega)=\int dt \; e^{-i \omega t} \langle \{ A^{+}(t), A(0) \} \rangle$ is the power spectrum  of the operator $A$ and $\langle . . . \rangle $ stands for the equilibrium (thermal) average.  The power spectrum $S_{X(Y)}(\omega)$ can be generally expressed in terms of the Fourier transform of the spin-spin correlator $C_{\alpha \beta}
(\vec{r}_{i},\vec{r}_{j}; t)=\langle \{ m_{\alpha}(\vec{r}_{i},t), m_{\beta}(\vec{r}_{j},0) \} \rangle$, with $\alpha, \beta=x,y,z$. Invoking the fluctuation-dissipation theorem~\cite{kubo1966}, we can write $C_{\alpha \beta}(k, \omega)=\coth (\beta \hbar  \omega/ 2) \chi''_{\alpha \beta}(k,\omega)$, where $\beta=1/k_{B}T$, with $k_{B}$ being the Boltzmann constant and $T$ the temperature.

For  the isotropic  bulk,  the spin-susceptibility tensor is diagonal, with $\chi_{xx}=\chi_{yy}$~\cite{Vignale2005,Flebus2018}. The response $\chi''_{xx}$ ($\chi''_{zz}$) stems from  spin fluctuations transverse (longitudinal) to the $\mathbf{z}$ axis, i.e., to the equilibrium orientation of the N\'eel order parameter in the bulk. As discussed in Ref.~\cite{Flebus2018},  transverse fluctuations of the spin density corresponds to one-magnon processes, i.e., the creation or annihilation of a magnon. The associated relaxation rate is proportional to the magnon spectral density  at the QI  resonance frequency. The latter is vanishing for $\omega \lesssim \Delta - 1/\tau_{s}$, with $\Delta$ being the spin-wave bulk gap. Thus, by tuning the QI frequency, one has $T^{-1}_{1}=0$. Furthermore, the imaginary part of the bulk transverse spin susceptibility scales as $\chi_{xx}''(\omega) \propto \omega^3$~\cite{FlebusOchoa2018}, leading to a vanishing dephasing rate, i.e., $T^{-1}_{2}=0$. 

The bulk longitudinal spin fluctuations correspond instead to two-magnon processes. The associated QI decoherence rate reflects the likelihood of magnons scattering  with energy  gain (loss) equal to the QI  frequency;  it is thus maximized at zero frequency, to then decrease with increasing QI frequency~\cite{FlebusOchoa2018}. However, magnons freeze out as the temperature drops below the spin-wave gap $\Delta$ and, by setting the temperature far below the bulk spin-wave gap,  $S_{X}(0)$ can be neglected.

\begin{figure}[t!]
\centering
\includegraphics[width=0.7\linewidth]{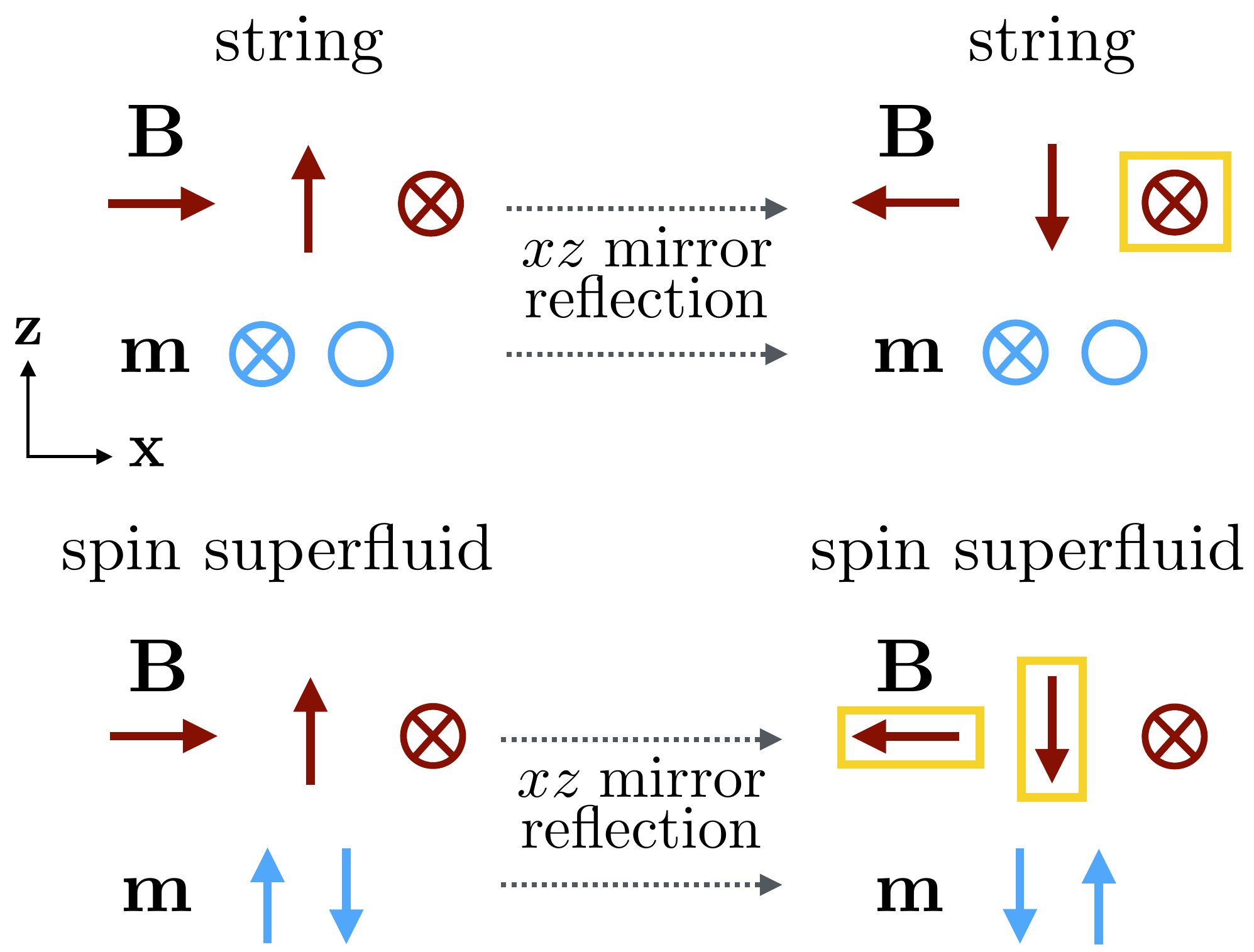}
\caption{Stray field $\mathbf{B}$ produced by a Bloch domain wall.  The yellow contour highlights the stray-field components (red arrows)  allowed by symmetry. Upper side:  the spin density $\textbf{m}$ (blue arrow) associated with the string-mode dynamics is invariant under mirror reflections through the $xz$ plane. The associated stray field  is allowed to have only a $y$ component. Lower side: the spin density  $\textbf{m}$ (blue arrow) produced by the spin-superfluid dynamics flips sign under mirror reflections through the $xz$ plane. Therefore, the allowed stray-field components are oriented along the $\mathbf{x}$ and $\mathbf{z}$ axes.}
\label{Fig2}
\end{figure}

For $d \ll \lambda$, we can relate the spin susceptibility of the magnetic film to the one associated with the DW modes as $\chi_{\alpha \beta}(\vec{r}_{i}, \vec{r}_{j} ;\omega)=\chi_{\alpha \beta}(|x_{i}-x_{j}|,\omega) \delta(y_{i}) \delta(y_{j})$.  In a Bloch DW, the order parameter lies along the $\mathbf{x}$ axis. Thus, according to the constraint $\mathbf{m} \cdot \mathbf{l}=0$, there is no finite spin density component along the $x$ direction. A finite out-of-plane spin density  (per unit of length) is engendered by the spin-superfluid dynamics, while  string-mode fluctuations give rise a spin density (per unit of length) along the $\mathbf{y}$ axis. In linear response, the longitudinal and transverse spin fluctuations do not interfere and can be considered separately~\cite{FlebusOchoa2018}. Hence, the relevant response functions are the $yy$ and $zz$ components of the imaginary spin susceptibility.  Since  $\chi''_{yy, zz}(\omega) \propto \omega^3$,  we can set $S_{X}(0)=0$. As it can be deduced from the symmetry argument illustrated in Fig.~\ref{Fig2}, the stray field generated by the string mode is parallel to $\mathbf{y}$ axis. Thus, when the QI quantization axis is oriented along the $y$ direction, we have $T^{-1}_{1}=0$, and, consequently, $T^{-1}_{2}=0$. Conversely, for the spin-superfluid mode, the associated stray-field components are oriented along the $\mathbf{x}$ and $\mathbf{z}$ axes, as shown in Fig.~\ref{Fig2}. The corresponding  relaxation rate is given by Eq.~(\ref{68}). Since the spin-superfluid dynamics give rise to spin fluctuations  transverse  to the equilibrium orientation of the order parameter, the QI relaxation rate can be minimized by tuning the QI frequency below the spin-superfluid gap.

 \textit{Qubit-qubit coupling.} Assuming the QI coupling to the antiferromagnetic spin density to be much smaller than the QI resonance frequency,  we can derive the effective qubit-qubit interaction by applying the lowest-order Schrieffer-Wolff transformation~\cite{Bravyi2011} to the interaction Hamiltonian
 \begin{align}
 \mathcal{H}_{\text{int}}=- \frac{\hbar \tilde{\gamma}}{2} \sum_{i=1,2} \boldsymbol{\sigma}_{i} \cdot \mathcal{R}_{x}(\pi/2) \mathbf{B}(\vec{r}_{i}).
 \end{align}
The resulting single-qubit terms such as $J \sigma_{1}^{+} \sigma_{1}^{+}$ vanish in the spin-$1/2$ subspace, while terms of the type $J \sigma_{1}^{+} \sigma_{1}^{-}$ can be reabsorbed in the definition of the QI frequency. Furthermore, for  $J \ll \omega$, we can neglect terms acting in the subspace ($|{\uparrow \uparrow}\rangle$, $|{\downarrow \downarrow}\rangle$), e.g., $J \sigma_{1}^{+}\sigma_{2}^{+}$.  Focusing on the spin-superfluid mode, which does not couple longitudinally  to the QI's,  no terms involving the operator $\sigma_{\tilde{z},1(2)}$ appear. These considerations lead to the effective qubit-qubit coupling Hamiltonian~(\ref{177}). 

A controlled-NOT and arbitrary one-qubit gates suffice for defining a universal set of gates. For a QI spin, single-qubit operations can be performed by locally applying  resonant microwave fields. A controlled-NOT gate can be decomposed into two iSWAP gates. By rewriting Eq.~(\ref{177}) as 
\begin{align}
\mathcal{H}_{c}=J (\sigma^{+}_{1} \sigma^{-}_{2} + \sigma^{+}_{2} \sigma^{-}_{1})\,,
\end{align}
 an iSWAP gate can be implemented as $U_{\text{iswap}}=\text{exp}(-i \mathcal{H}_{c} t_{J}/\hbar)$, with $t_{J}=\pi/ 4 J$~\cite{Schuch2003}.

The  qubit-qubit interaction mediated by the transverse and longitudinal bulk spin waves can be neglected when the temperature and the QI frequency lie much below the bulk spin-wave gap.
For any QI frequency, instead, the string mode mediates an RKKY-like interaction between the qubits, which can be found as
 \begin{align}
 \mathcal{H}_{c}=\left( \frac{\gamma \tilde{\gamma}}{2}\right)^2\int   \frac{dk}{2\pi}\; \chi_{yy}'(k, 0) \cos (kL) \sigma_{ \tilde{z},1} \sigma_{ \tilde{z},2}\,.
 \end{align}
 In our model, the real part of the static spin susceptibility decays very rapidly, i.e., $\chi_{yy}'(x) \propto \delta(x)$. Accounting for a finite exchange stiffness $A_{s}$ associated to the spin density field, which translates into adding a term   $\propto A_{s} (\vec{\nabla} \textbf{m})^2$ to Eq.~(\ref{92}), would introduce a finite decay length $\lambda_{s}=\sqrt{A_{s}/K}$. However, we can anticipate this length to be on the atomistic scale, thus much shorter than the characteristic lengthscale $\ell_{s}$  that controls the strength of the qubit-qubit coupling mediated by the spin-superfluid~(\ref{177}).
 
\textit{Estimate.}
As QI prototype we consider a NV center, i.e., a spin triplet with an intrinsic dephasing time of $T_{2}\sim 100$ ms at temperatures of few Kelvins~\cite{BarGill2013}.  By tuning the magnetic field, we can isolate a subsystem of the spin triplet and treat a NV center as an effective two-level system.   
Writing $A=J S^2$ and $\chi=\hbar^2/8 J S^2 a^2$ \cite{Auerbach}, with $J$ being the Heisenberg exchange, $S$ the spin and $a$ the lattice constant, we set $S\approx 1$ and $a \approx 5 \; \dot{\text{A}}$. We take $\gamma = \tilde{\gamma} \approx 2 \mu_{B}/ \hbar$, with $\mu_{B}$ being the Bohr magneton, $\Delta_{s} \approx 1$ GHz and  $\lambda \approx 10$ nm. For $\ell \sim 1 \; \mu\text{m}$,  we obtain, for a magnon mode with $k \sim 1/d$,  a coupling strength $g \sim 10 \; $kHz. We note that the latter, for a given DW width, does not depend on the exchange stiffness.
From the LLG phenomenology~\cite{FlebusOchoa2018}, we have $\tau_{s}=2 \chi / s \alpha$, where $\alpha$ is the Gilbert damping, which we set to $\alpha \sim 10^{-4}$. For $J\sim 0.1-1$ THz, we find a cooperativity $C \sim 10\div100$, much higher than the one associated with, e.g., hybrid  devices based on NV centers and mechanical resonators~\cite{Kolkowitz2012}. Plugging Eq.~(\ref{susceptibilitySS}) into Eq.~(\ref{177}) and setting $ \Delta_{s}-\omega \sim 1$ MHz,  we find, for $J \sim 0.1$ THz, an operation rate  $t_{J}^{-1}\sim 10 \; $kHz for $L \sim 1$ $\mu$m. 
The QI relaxation rate induced by the spin-superfluid noise is of order of tens of Hz at $T=100$ mK, i.e., two orders of magnitude smaller than the operation rate. Setting  $J \sim 1$ THz  decreases the  operation rate to $t_{J}^{-1}\sim 1 \; $kHz, but it leads to a ratio between the latter and the QI relaxation rate of the order $\sim 10^4$.

\textit{Discussion.}  Recently, strongly-localized quantized magnetic solitons with nonlinear features have been  proposed  as carriers of quantum information~\cite{Takei2018}. Here, changing the perspective, we focus on the soft bosonic modes of extended domain walls to mediate coupling between magnetic qubits that are extrinsic to the antiferromagnetic medium.
Specifically,  we show that  the spin-superfluid mode harbored by an antiferromagnetic DW can mediate a tunable coherent coupling between spin qubits separated on a micron scale, i.e., a distance larger than, e.g., the one required to address NV centers separately \cite{diffractionlimit}. We propose a universal set of gates that can be  switched on and off via an external magnetic field.

Future works should more systematically address the role of quenched disorder, the effects of higher-order magnon processes associated with the DW dynamics, and the decoherence from other sources. These may include the phononic background and dynamic spin impurities in the magnetic medium (which go beyond the Gilbert-damping phenomenology of collective dissipation).

The authors thank T. van der Sar and C. Du for helpful discussions. B.F. has been supported by the Dutch Science Foundation (NWO) through a Rubicon grant and Y.T. by NSF under Grant No. DMR-1742928.


\begin{thebibliography}{99}
\bibitem{diffractionlimit} Generally, the spatial separation between two NV centers should be larger than the wavelength $\lambda \sim 1/2$~$\mu$m of the laser deployed for their initialization and read-out.
\end{thebibliography}
\end{document}